\documentclass[reprint]{revtex4-1}

\usepackage{graphicx}
\usepackage{amsmath, amsfonts, amssymb}

\begin{document}

\title{Magnetic Moments of Lanthanide van der Waals  Dimers}

\author{Joseph McCann$^{1}$}
\author{John L. Bohn$^{1}$}
\author{Lucie D. Augustovi\v{c}ov\'{a}$^{2}$}
\affiliation{$^{1}$JILA, NIST, and Department of Physics, University of Colorado, Boulder, CO 80309-0440, USA}
\affiliation{$^{2}$Charles University, Faculty of Mathematics and Physics, Department of Chemical Physics and Optics, Ke Karlovu 3, CZ-12116 Prague 2, Czech Republic}

\date{\today}

\begin{abstract}
Loosely bound van der Waals dimers of lanthanide atoms, as might be obtained in ultracold atom experiments,  are investigated.  These molecules are known to exhibit a degree of quantum chaos, due to the strong anisotropic mixing of their angular spin and rotation degrees of freedom.  Within a model of these molecules, we identify different realms of this anisotropic mixing, depending on whether the spin, the rotation, or both, are significantly mixed by the anisotropy.  These realms are in turn generally correlated with the resulting magnetic moments of the states.  
\end{abstract}

\maketitle

\section{Introduction: Molecular Complexity}

The harmony and quietude of ultracold atoms were shattered in 2014, with the discovery that collisions of erbium atoms at 300 nK exhibited the telltale signs of quantum chaos \cite{Frisch14_Nat}, followed  by similar observations in dysprosium \cite{Baumann14_PRA,Maier15_PRX} and thulium \cite{Khlebnikov19_PRA}, as well as dysprosium-erbium mixtures \cite{Durastante20_PRA}.  Magnetic field scans revealed not only an unprecedented host of Fano-Feshbach resonances, but also that the magnetic field locations of these resonances appeared to be  distributed according to the predictions of random matrix theory, a finding suggestive of quantum chaos.  The observations attest to the unexpected complexity of weakly-bound lanthanide dimer molecules within several GHz of their dissociation threshold.  

This observation is a challenging one to interpret within the paradigms of quantum chaos.  For one thing, it is not completely clear that the resolution in the experiments was capable of detecting all the resonances.  Disregarding the very narrow ones may bias the spectrum toward revealing less chaos than it truly possesses \cite{MurPetit15_PRE}.  Further, while a common statistical analysis tool -- the Brody function that characterizes  the distribution of nearest-neighbor spacings  in the spectrum -- showed evidence of chaos, its appropriateness has been  called into question.  The claim was made that perhaps the data were better fit by a semi-Poisson distribution, revealing the system to be quasi-integrable \cite{Roy17_EL}.   Another complication arises in the unusual nature of the experiments.  Typically, quantum chaos studies the spectrum and eigenfunctions of a given Hamiltonian.  By contrast, the ultracold experiments determine a spectrum of magnetic field values at which Fano-Feshbach resonances occur.  Thus the spectrum identifies  a single, zero-energy eigenstate from each of an ensemble of different Hamiltonians, one for each magnetic field.  It is not immediately obvious how to view nearest-neighbor spacings in such a circumstance \cite{Augustovicova18_PRA}.

These issues were clarified by a  combination of experiment and theory that looked at the Fano-Feshbach spectra in both Er and Dy, comparing them to the results of large-scale scattering computations \cite{Maier15_PRX}.  It was concluded that the observed Brody parameters corresponding to the magnetic field spectra could indeed be accounted for.  This required appreciating that, not only were the atoms complex, with many internal spin states, but also that these states were strongly coupled by anisotropic interactions during their collision \cite{Kotochigova11_PCCP}.  More significantly, the calculation revealed that the  degree to which the energy spectrum is chaotic is contingent on the value of the magnetic field at which the spectrum is generated.   Further evidence of order amid the chaos appears in measurements of molecular bound states by magnetic field modulation spectroscopy near broad Fano-Feshbach resonances \cite{Maier15_PRA,Lucioni18_PRA}.  These measurements and subsequent analysis identify these bound states as being of essentially single-channel, $s$-wave character, unencumbered by significant coupling to other angular momentum states.  

Numerical models of the spectrum can of course reveal information about the molecules that the experiment is not yet privy to.  Thus Ref.~\cite{Makrides18_Sci} applied more sophisticated statistical tests, including a family of information-theoretic entropies, to the numerical spectrum from Ref.~\cite{Maier15_PRX}, looking at the energy range 0.5 GHz below threshold.  Unlike  the experiments, this calculation was able to evaluate a significant region of the energy spectrum at any desired magnetic field, along with the energy eigenstates.  A main conclusion is that the spectrum exhibits ``multifractal'' behavior,  quantifying  the degree to which  the spectrum appears chaotic.  Moreover, various measures of quantum chaos were seen to increase as the magnetic field grew larger, and channel mixing increased.

Generally, analysis in terms of entropies is useful in locating complex systems along the complexity scale.  For truly random systems, whose spectra are well-described statistically in random matrix theory, this theory provides upper limits to the values of entropies \cite{Kota_book}.  Many physical systems fall short of this upper limit, and have sub-maximal entropies, suggesting that some order remains \cite{Kota_book,dAlessio16_AP}; this is the case for the model  of Ref.~\cite{Makrides18_Sci}.  A consequence of the previous analyses is therefore that lanthanide dimers just below their dissociation threshold are complicated, yes, but they are not thoroughly chaotic.  There may be something orderly about them, at least in some eigenstates, that can be expressed in terms of the familiar ingredients of molecular physics, namely, vibrations, rotations, and spins, and their quantum numbers.  In favorable cases, this would entail identifying good quantum  numbers, or nearly good quantum  numbers, so that an appropriate Hund's case might be identified.  On the other hand, it may be the case that in some states this is not possible, and the objects of quantum chaos theory become the appropriate tools for describing the states, or at least specific ensembles of states.  

 In this spirit, this article will examine weakly bound van der Waals Dy$_2$ dimers, and seek to understand which, if any, quantum  numbers remain good, and under what circumstances.  To give a specific context, we associate this search to concrete observable quantities of the molecular states, namely, their magnetic moments.  These weakly bound dimers can be created in the laboratory, say by magnetoassociation, and their magnetic dipole moments can be measured \cite{Frisch15_PRL}.  They may also perhaps be accessible by microwave spectroscopy.  In any event, we pursue statistical aspects of a distribution of magnetic moments, which in general are uncorrelated to the energies of the states, but which provide insight into the composition of molecular wave functions.  
 
 In the current article we focus for simplicity on  molecules in zero magnetic field, where the total angular momentum is conserved.  We find that the states with fairly well-defined angular momentum quantum numbers tend to be those that are stretched, with near-maximal values of certain angular momenta.  As a consequence, the more orderly versus more chaotic states can be distinguished, on average, by the values of their  magnetic moments.

\section{Scope of the Paper}

We contemplate a diatomic molecule composed of two identical lanthanide atoms, such as Dy, Er, Tm, etc.  In practice, for the calculations below, we will use Dy, and explicitly consider bosonic isotopes with zero nuclear spin.  Each of the atoms  has spin ${\bf j}$ and magnetic moment  $\boldsymbol{\mu}$
 $= - g \mu_B {\bf j}$, where $\mu_B = e \hbar / 2 m_e c$ is the Bohr magneton, expressed in cgs units; and $g$ is the atom's $g$-factor, here defined as a positive number.  Thus in an applied magnetic field that defines the laboratory $z$ axis, the energies of the atom depend on the field as $g \mu_B B m$, for state $|jm \rangle$.  This expression for energy shift is accurate for fields small enough such that $j$ remains a good quantum number.  

When two lanthanide atoms are combined into a van der Waals dimer, the resulting molecule is a composite object with total angular momentum ${\bf J} = {\bf j}_1 + {\bf j}_2 + {\bf L}$, where ${\bf j}_i$ is the spin of the $i$th atom and ${\bf L}$ is the orbital angular momentum of the atoms about their center of mass.  We use ${\bf L}$ for this quantity to draw the analogy with the partial wave angular momentum in ultracold collisions of the atoms.  Since the atoms are electrically neutral, the orbital motion does not contribute to the magnetic moment of the molecule, whose moment is therefore $\boldsymbol{\mu} = -g {\bf j}_1 - g{\bf j}_2$.  The extreme values of the moment occur when $m_1=m_2= \pm j$, whereby the molecular magnetic moment must lie between $\pm g \mu_B (2j)$.  For Dy, with $j=8$ and $g=1.2416$, these bounds are $\pm 19.9 \mu_B$.  Various states of the molecule will have magnetic moments between these two limits, depending on the details of how the state is constructed.  The moment is therefore a probe of how the separate angular momenta work together in the molecule.   

The number of possible energy eigenstates of these molecules is vast.  To lend focus to the current investigation, we strongly constrain its scope, to a set of molecular states close to experimental reality.  Specifically, we will consider a pair of spin-stretched Dy atoms initially in their lowest-energy state $|jm \rangle=|8-8 \rangle$.  A small field may  be applied to remove the degeneracy of the states.  These atoms are assumed to collide via an $s$-wave collision with $L=0$, hence the total angular momentum of the atom pair is $|JM \rangle = |16-16 \rangle$, and this angular momentum is considered to be conserved in sufficiently small magnetic field.  It is conceivable, if not necessarily easy, to perform a microwave spectroscopy experiment that associates the free atoms into weakly-bound states of the Dy$_2$ dimer.  Measurements of the energy $E_{\alpha}$ at two distinct, small magnetic fields allow the determination of the magnetic moment of state $|\alpha \rangle$ via $\mu_{\alpha} = \Delta E_{\alpha} / \Delta B$.  The statistical distribution of the moments so defined are the subject of our inquiry.

\section{Model}

Given the complexity of the lanthanide van der Waals dimers, a complete and accurate {\it ab initio} theory of their structure remains challenging.  Nevertheless, guided by previous models, some of the salient features of the dimers are apparent.  The predominant features of the interatomic interactions at large interatomic separation consist  of the magnetic dipole-dipole interaction and the anisotropic van der Waals interaction, the latter of which is believed to be primarily responsible for the channel mixing that generates some degree of chaos in the molecular spectrum \cite{Maier15_PRX}.  We will therefore include these interactions in some detail, with an emphasis on their representation in alternative angular momentum coupling schemes.    By contrast, our representation of the Born-Oppenheimer potentials will be somewhat more schematic, as less important to the present analysis.  

\subsection{Basis Sets}

To attain the total angular momentum $J$ requires coupling the spins of the two atoms to their relative orbital angular momentum.  Formally, this can be done in either the lab frame, or else in the body frame of the molecule.  In either case we use as an intermediate the total spin angular momentum of the two atoms, ${\bf j}_{12} = {\bf j}_1 + {\bf j}_2$.  We  contemplate two  basis sets for the molecule, resembling the Hund's cases (a) and (b) familiar from the theory of diatomic molecules.  These basis sets are as follows:

{\it Body-fixed frame (BF):}  The individual spins ${\bf j}_1$, ${\bf j}_2$ are coupled to a total spin ${\bf j}_{12}$, with projection $\Omega$ along the intermolecular (body frame) axis.  The total angular momentum is $J$ and its projection on the lab axis is $M$.  Suppressing the notation $j_1$ and $j_2$,  this basis is
\begin{eqnarray}
|j_{12} \Omega; JM \rangle_u = |j_{12} \Omega \rangle |\Omega JM \rangle,
\label{eq:body_basis}
\end{eqnarray}
where
\begin{eqnarray}
|j_{12} \Omega \rangle &=& \sum_{\omega_1 \omega_2} |j_1 \omega_1 \rangle |j_2 \omega_2 \rangle
\langle j_1 \omega_1 j_2 \omega_2 | j_{12} \Omega \rangle, \\\
|\Omega JM \rangle &=& \sqrt{ \frac{ 2J+1 }{ 8\pi^2 } }D^{J*}_{M\Omega}(\phi,\theta,\gamma),
\end{eqnarray}
with $\omega_i$ the projection of spin ${\bf j}_i$ on the interatomic axis.  
Here the Wigner rotation matrix $D$ is a function of the Euler angles $(\phi,\theta,\gamma)$ that relate the body frame to the lab frame.  We include the subscript $u$ for ``unsymmetrized,'' so that we don't need to carry around an extra subscript for the symmetrized version below.  The spins of the atoms are quantized along the body-frame axis, identifying this basis set as analogous to Hund's case (a).

Symmetrized according to even exchange of identical bosons, the basis set becomes 
\begin{widetext}
\begin{eqnarray}
|j_{12} {\bar \Omega}; JM \rangle = \frac{ 1 }{ \sqrt{ 2(1 + \delta_{{\bar \Omega}0}) } }
\left[ |j_{12}{\bar \Omega}; JM \rangle_u + (-1)^J |j_{12} - {\bar \Omega};JM \rangle_u \right].
\end{eqnarray}
\end{widetext}
Here ${\bar \Omega} = |\Omega|$ is intrinsically non-negative and takes the values ${\bar \Omega} = 0,1, \dots, \min(j_{12},J)$.  Note that ${\bar \Omega}=0$ is possible only when $J$ is even.

{\it Coupled lab-frame (CLF):} The individual spins ${\bf j}_1$, ${\bf j}_2$ are coupled to a total spin ${\bf j}_{12}$, with projection $m_{12}$ on the lab axis:
 \begin{eqnarray}
 |j_{12}m_{12} \rangle = \sum_{m_1  m_2}  |j_1 m_1 \rangle |j_2 m_2 \rangle
  \langle  j_1 m_1 j_2  m_2 | j_{12}m_{12} \rangle. \nonumber  \\
  \end{eqnarray}
   The rotation of the molecule is described by the orbital angular momentum $L$ and its lab projections $M_L$ of the atoms about their center of mass, with wave function
   \begin{eqnarray}
   |L M_L \rangle = Y_{L M_L}(\theta,\phi) = \sqrt{ \frac{ 2L+1 }{ 4 \pi } } C_{L M_L}(\theta, \phi), \nonumber \\
   \end{eqnarray}
   where $C_{L M_L}$ is a reduced spherical harmonic.  These are coupled into the total angular momentum $J$ with lab projection $M$:
 \begin{eqnarray}
 |[j_{12} L]JM \rangle &=& \sum_{m_{12} M_L} |j_{12} m_{12} \rangle |LM_L \rangle
 \langle j_{12} m_{12} L M_L | JM \rangle. \nonumber \\
 \label{eq:lab_basis}
 \end{eqnarray}
 This basis is already symmetric under exchange of the identical bosons, provided that $j_{12}+L$ is  even.  Quantization of the atomic spins in the laboratory frame identifies this basis set as analogous to Hund's case (b).  
 
The two basis sets are related by a unitary transformation with matrix elements
\begin{widetext}
 \begin{eqnarray}
\langle j_{12} {\bar \Omega}; JM | [j_{12}^{\prime} L^{\prime} ] J^{\prime} M^{\prime} \rangle
= \frac{ 2 }{ \sqrt{ 2(1+\delta_{{\bar \Omega}0}) } }
(-1)^{M-{\bar \Omega}} \sqrt{ 2L^{\prime}+1} 
\left( \begin{array}{ccc} j_{12} & L^{\prime} & J \\ {\bar \Omega} & 0 & - {\bar \Omega} \end{array} \right)
\delta_{j_{12} j_{12}^{\prime}} \delta_{JJ^{\prime}} \delta_{MM^{\prime}}.
\end{eqnarray}
\end{widetext}
 Thus the various pieces of the Hamiltonian can be cast in either basis, as convenient, and easily transformed to the other as necessary.  
 
 For the examples considered in this paper,  linked to a presumed  initial state defined by $s$-wave scattering with $L=0$, our identical bosons can only access states with even values of $j_{12}$, that is, {\it gerade} states of the interatomic potential energy surfaces.  We will impose this restriction on the results below.

 \subsection{Hamiltonian}
 
 The Hamiltonian can be written as a sum of contributions,
\begin{eqnarray}
H = T + H_\mathrm{BO} + H_\mathrm{dd} + H_\mathrm{ad} + H_\mathrm{B},
\end{eqnarray}
which are, in order: kinetic energy; the Born-Oppenheimer potentials responsible primarily for short-range interactions; the long-range dipole-dipole interaction; the long-range anisotropic dispersion interaction; and the magnetic field Hamiltonian.  Bearing in mind the transformations between the basis sets, different parts of the Hamiltonian are easy to write in different bases and transformed to the other as necessary.  Thus the interaction terms $H_{\mathrm BO}$, $H_{\mathrm dd}$, and $H_{\mathrm ad}$ are easily written in the body frame, while $T$ and $H_{\mathrm B}$ take simple forms in the lab frame.

\subsubsection{Kinetic Energy}

In the usual way, we write the total wave function in the form $\Psi(R,\sigma) = R^{-1}f(R,\sigma)$ where $\sigma$ denotes all coordinates other than the interatomic spacing $R$.  In this case the kinetic energy amounts to a radial component and a centrifugal component,
\begin{eqnarray}
T = - \frac{ \hbar^2 }{ 2 m_r } \frac{ d^2 }{ dR^2 } + T_\mathrm{cent},
\end{eqnarray}
where $m_r$ is the reduced mass of the colliding pair, and the centrifugal part is diagonal in the laboratory basis,
\begin{eqnarray}
T_\mathrm{cent} = - \frac{ \hbar^2 L(L+1) }{ 2m_r R^2 } \delta_{j_{12}j_{12}^{\prime}} 
\delta_{LL^{\prime}} \delta_{JJ^{\prime}} \delta_{MM^{\prime}}.
\end{eqnarray}

\subsubsection{Born-Oppenheimer potentials}

The Born-Oppenheimer part is determined, in general, from detailed electronic structure calculations.  These have been carried out for Dy and Er, but instead we find it convenient to use simpler, analytic forms as a stand-in for these potentials.  

The molecular axis is an axis of rotational symmetry for the interactions among the electrons and nuclei that make up the molecule, whereby ${\bar \Omega}$ is a good quantum number, and the body frame basis makes sense.  The total spin angular momentum $j_{12}$ need not be a good quantum number and different values may be  somehow coupled together.  However, the parity of $j_{12}$ is good: the {\it gerade} states have even $j_{12}$ values, and the {\it ungerade} states have odd $j_{12}$ values.  This Hamiltonian is moreover independent of the total rotational state of the molecule, hence independent of $J$ and $M$.  

We simplify diagonal elements of $H_{\mathrm BO}$ by employing a set of Lennard-Jones potentials,
\begin{eqnarray}
\langle j_{12} {\bar \Omega}; JM | H_{\mathrm BO} | j_{12} {\bar \Omega}; JM \rangle
= \frac{ C_{12}({\bar \Omega}, j_{12}) }{ R^{12} } - \frac{ C_6 }{ R^6 }. \nonumber \\
\end{eqnarray}
Here each channel as assumed to have the same isotropic van der Waals coefficient $C_6$, as the anisotropy is dealt with separately.  Each diagonal channel may have a different $C_{12}$ coefficient, which may be drawn from a statistical ensemble so that these potentials have random scattering lengths, if desired.  However, in the results below, we employ a particular value of $C_{12}$ in all channels.   Likewise, it would be possible to generate random matrix elements that couple different values of $j_{12}$, but we have not done so here.  This approach exploits the observation of Ref.~\cite{Maier15_PRX} that the dominant channel coupling occurs due to the anisotropic van der Waals interaction.  For the calculations described below, we use $C_6 = 2274$ au \cite{Lepers_private}, and artificially truncate the potentials at small $R$ using the same value $C_{12} = 1.0 \times 10^{11}$ in all channels.  

\subsubsection{Dipole-dipole Interaction}

The dipole-dipole interaction is also naturally described in the body frame, where it is diagonal in ${\bar \Omega}$,
\begin{widetext}
\begin{eqnarray}
\langle j_{12} {\bar \Omega}; JM | V_{dd} | j_{12}^{\prime} {\bar \Omega}^{\prime}; 
J^{\prime}M^{\prime} \rangle
&=& - \frac{ \sqrt{30} (g \mu_B )^2 }{ R^3 } \frac{ 1 + (-1)^{j_{12}+j_{12}^{\prime}} }{ 2 }  
\delta_{{\bar \Omega} {\bar \Omega}^{\prime}} \delta_{JJ^{\prime}} \delta_{MM^{\prime}} \\
&& \times (-1)^{j_{12}-{\bar \Omega}} j(j+1)(2j+1) \sqrt{ (2j_{12}+1)(2j_{12}^{\prime}+1) }
\nonumber \\
&&\times \left\{ \begin{array}{ccc} j_{12} & j_{12}^{\prime} & 2 \\ j & j & 1 \\ j & j & 1 \end{array} \right\}
\left( \begin{array}{ccc} j_{12} & 2 & j_{12}^{\prime} \\ {\bar \Omega} & 0 & - {\bar \Omega} \end{array} \right), \nonumber
\end{eqnarray}
\end{widetext}
where $j=j_1=j_2$.  The coupled spin $j_{12}$ is not conserved by this interaction, but its parity is.

\subsubsection{Anisotropic Dispersion Interaction}

For large $R$, the atoms also exert anisotropic dispersion forces on each other.  These are evaluated in detail in Ref.~\cite{Lepers_chapter}.  The dominant dispersion term is of course the isotropic one described above.  Second to this, and the only non-negligible correction in this context, is the term given explicitly in the uncoupled, body frame basis by \cite{Lepers_chapter}
\begin{widetext}
\begin{eqnarray}
\langle j_1 \omega_1 j_2 \omega_2 | V_{ad} | j_1 \omega_1 j_2 \omega_2 \rangle
= \sqrt{ \frac{ 5 }{ 2 } } \frac{ C_{6,20} }{ R^6 }
\Big( \langle j_1 \omega_ 2 20 | j_1 \omega_1 \rangle +\langle j_1\omega_2 20 | j_2 \omega_2 \rangle \Big),
\end{eqnarray}
\end{widetext}
where $C_{6,20}$ is a numerical coefficient derived in perturbation theory.  Adapting this to the coupled body frame basis gives the matrix elements
\begin{widetext}
\begin{eqnarray}
\langle j_{12} {\bar \Omega}; JM | V_{ad} | j_{12}^{\prime} {\bar \Omega}^{\prime}; J^{\prime} M^{\prime} \rangle
&=& \frac{ C_{6,20} }{ R^6 }  \frac{ 1 + (-1)^{j_{12}+j_{12}^{\prime}} }{ 2 } 
\delta_{{\bar \Omega} {\bar \Omega}^{\prime}} \delta_{JJ^{\prime}} \delta_{MM^{\prime}} \nonumber \\
&& \times (-1)^{2j-{\bar \Omega} } \sqrt{ 10 } \sqrt{ (2j+1)(2j_{12}+1)(2j_{12}^{\prime}+1) }
\nonumber \\
&& \times \left\{ \begin{array}{ccc} j & j_{12} & j \\ j_{12}^{\prime} & j & 2 \end{array} \right\}
\left( \begin{array}{ccc} j_{12} & 2 & j_{12}^{\prime} \\ {\bar \Omega} & 0 &  -{\bar \Omega} \end{array} \right).
\label{eq:Vad_matrix}
\end{eqnarray}
\end{widetext}
  Just as for dipoles, ${\bar \Omega}$ is conserved, as is the parity of $j_{12}$, but $j_{12}$ itself is not.  
  
  The constant $C_{6,20}$ in front of this expression is subject to considerable uncertainly in the literature.  In general, the strength of the anisotropic dispersion contribution is characterized by diagonalizing the matrix  (\ref{eq:Vad_matrix}), exclusive of $1/R^6$, and defining $\Delta C_6$ as the difference between the maximum and minimum eigenvalues.  Reported values of this constant include $\Delta C_6 = 5.8$ au \cite{Li17_JPB}, $\Delta C_6 = 14$ au \cite{Lepers_private}, $\Delta C_6 =  25$ au \cite{Kotochigova11_PCCP}, and  $\Delta C_6 = 174$ au \cite{Makrides18_Sci}.   In the interest of incorporating significant channel mixing in the model, we will use the last value, which corresponds to $C_{6,20}=-44.4$ au.

\subsubsection{Magnetic Field Hamiltonian}

The magnetic field acts on the magnetic moments of the atoms separately,
\begin{eqnarray}
H_B =  g \mu_B B T_0^1({\bf j}_1) + g \mu_B B T_0^1({\bf j}_2).
\end{eqnarray}
Its matrix elements are conveniently written in the coupled laboratory frame as
\begin{widetext}
\begin{eqnarray}
\langle [j_{12}L] JM | H_B | [j_{12}^{\prime}L^{\prime} ] J^{\prime} M^{\prime} \rangle
&=&  g \mu_B B \left[ (-1)^{j_{12}} + (-1)^{ j_{12}^{\prime}} \right] \delta_{LL^{\prime}} \delta_{MM^{\prime}}
(-1)^{J-M+J^{\prime}+j_{12}+L} \\
&& \times \sqrt{ j(j+1)(2j+1)(2j_{12}+1)(2j_{12}^{\prime}+1)(2J+1)(2J^{\prime}+1) } \nonumber \\
&& \times \left\{ \begin{array}{ccc} j_{12}& J & L \\ J^{\prime} & j_{12}^{\prime} & 1 \end{array} \right\}
\left\{ \begin{array}{ccc} j & j_{12} & j \\ j_{12}^{\prime} & j & 1 \end{array} \right\}
\left( \begin{array}{ccc} J & 1 & J^{\prime} \\ -M & 0 & M \end{array} \right). \nonumber
\end{eqnarray}
\end{widetext}
  This interaction is capable of mixing different values of the total angular momentum that differ by 1.  $j_{12}$ could also change by 1, except that its parity must be conserved.  Therefore this matrix element is diagonal in $j_{12}$.  
  
  \subsection{Vibration}
  
 Each basis sets above defines a particular realization of a set of $R$-dependent diabatic channels, which would be suitable for scattering calculations.    We denote for brevity this set of quantum numbers by the collective ket $|d \rangle $, which stands for either the body frame channel basis (\ref{eq:body_basis}) or else the lab frame channel basis (\ref{eq:lab_basis}).  The wave function $R\Psi$ is acted upon by  the Hamiltonian
    \begin{eqnarray}
  H = - \frac{ \hbar^2 }{ 2m_r } \frac{ d^2 }{ dR^2 } + V_\mathrm{d}(R) + V_\mathrm{od}(R),
  \end{eqnarray}
  where $V_\mathrm{d}$ is a set of diabatic potential curves, consisting of the diagonal matrix elements of the Hamiltonian $T_\mathrm{cent} + V_\mathrm{BO} + V_\mathrm{dd} + V_\mathrm{ad} + H_\mathrm{B}$ as expressed in this basis, while $V_\mathrm{od}$ contains all of the off-diagonal matrix elements.
  
  Each potential $V_\mathrm{d}$ possesses a set of vibrational bound states, given by
  \begin{eqnarray}
  - \frac{ \hbar^2 }{ 2 m_r } \frac{ d^2 f_{d,v_d} }{ dR^2 }+ V_\mathrm{d} f_{d,v_d} = E_{d,v_d} f_{d,v_d}.
  \end{eqnarray}
 The set of states
  \begin{eqnarray}
  | i \rangle \equiv |d,v_d \rangle = |d \rangle f_{d,v_d}
  \label{eq:ro_vibrational_basis}
  \end{eqnarray}
  therefore constitute an approximate set of molecular states for our lanthanide diatom.  These states represent the molecular states as accurately as possible, while still retaining rigorously good values of the angular momentum quantum numbers $d$ of the body- or lab-frame, and a well-defined vibrational quantum number.   We will refer to these as the molecular basis states.  If they are minimally mixed, then their quantum  numbers are still a valid way to express the states of the molecule;  if they are strongly mixed, then they serve to identify what, exactly, is being mixed on the way toward making the molecule chaotic.  
  
  The wave function can then be expanded in this basis,
  \begin{eqnarray}
  f = \sum_{d v_d} c_{d v_d} |d v_d \rangle.
  \end{eqnarray} 
  Solving the Schr\"odinger equation amounts to diagonalizing the Hamiltonian ${\cal H}$ in the extended basis $|d v_d \rangle$.  
  The diagonal elements of this matrix are
  \begin{eqnarray}
  \langle d v_d | {\cal H} | d v_d \rangle = E_{d v_d},
  \end{eqnarray}
 while those matrix elements explicitly off-diagonal in $d$ are given by
  \begin{eqnarray}
  \langle d v_d | {\cal H}  | d^{\prime} v_d^{\prime} \rangle = 
   \int dR f_{dv_d}(R) \langle d | V_\mathrm{od} | d^{\prime} \rangle f_{d^{\prime}v_d^{\prime}}(R) \nonumber \\
   \end{eqnarray}
   and matrix elements of $V_\mathrm{od}$ can be computed term by term, knowing the explicit form of the various terms as given above.  
   
   We obtain the molecular spectrum by diagonalizing the Hamiltonian matrix in these terms.  The vibrational states are computed in each diabatic channel by a Fourier grid Hamiltonian method \cite{Marston89_JCP}, subject to box boundary conditions at the radius $R_2=400 a_0$.  This truncation may alter those states within about $\hbar^2/(2m_r R_2^2) \approx 0.1$ MHz, which represents a negligible part of the spectrum we study.  In order to achieve convergence of the final spectrum, we must include states in the quasi-continuum of this box.

\subsection{Magnetic moments of the basis states: $g$-factors}

The basis states defined in the previous section afford the simplest model of the magnetic moment distribution of the molecules.  In the absence of channel coupling, and in the limit of zero magnetic field, all the quantum numbers remain good.  In this case, the magnetic moments of the states may be described by analytical formulas.

  We can write the magnetic moments in terms of $g$-factors as 
\begin{align}
\mu_{i}^\mathrm{basis} = g(i) \mu_B M.
\label{eq:g_def}
\end{align}
Expressions for the $g$-factors can then be derived by evaluating diagonal matrix elements of the magnetic Hamiltonian.   In the body frame these are
 \begin{align}
g(j_{12},{\bar \Omega},J) =   \frac{ {\bar \Omega}^2  }{ J(J+1) },
 \label{eq:statistical_body}
 \end{align}
 while in the lab frame they are
 \begin{align}
g(j_{12},L,J) =  \frac{ 1 } { 2 } \left[ 1 + \frac{ j_{12}(j_{12}+1) - L(L+1) }{ J(J+1) } \right].
\label{eq:statistical_lab}
\end{align}
These are, of course, familiar expressions in molecular physics \cite{BC}.  

One can then define a statistical distribution of magnetic moments for either of these forms, by simply giving the occurrence of each possible quantum number equal weight.  The statistical distribution of the moments in the body frame would count each value of $j_{12}$ from $0$ to $2j$, counting only even values for the {\it gerade} states we consider here; and values of ${\bar \Omega}$ from $j_{12}$ up to $2j$.  For Dy with $j=8$, this amounts to 81 possibilities.  Counting each such possibility equally would give a distribution with the mean and standard deviation for the magnetic moments
\begin{align}
{\bar \mu}_{\mathrm body} &=   g \mu_B M \frac{ 2j+3 }{ 12(j+1) }   \\
\sigma( \mu )_{\mathrm body}&= g \mu_B |M| 
\sqrt{ \frac{ (28j^2 + 24j -9) (j+2)(2j+3) } {    720j^2 (2j+1) } } \nonumber
\label{eq:mean_moment_body}
\end{align}
Likewise, in the lab frame the quantum number for {\it gerade} states will run even values of  $j_{12}$ from $0$ to $2j$, while $L$ runs, also in even values, from $|J-j_{12}|$ to $J+j_{12}$, where $J=2j$ in the examples considered here.  Counting each possibility equally gives, for the lab frame,
\begin{align}
{\bar \mu}_{\mathrm lab} &= g \mu_B M  \frac{ 2j+3 }{ 12(j+1) } \\
\sigma( \mu )_{\mathrm lab} &= 
g \mu_B |M| \sqrt{  \frac{ 264j^3 + 548j^2 + 286j-3 }{ 720 j^2(2j+1) } }  \nonumber
\label{eq:mean_moment_lab}
\end{align}
The mean value of the magnetic moment is the same in either basis \cite{mean_note},  but the standard deviations are quite different.  

For the $j=8$ Dy atom in our examples, and in the state where $J=16$, $M=-16$, we find that, in the body frame, the mean of the magnetic moment distribution is $-3.49 \mu_B$, and the standard deviation of the distribution is $4.35 \mu_B$.  By contrast, in the body frame the mean of the distribution is the same, but its standard deviation is $9.32 \mu_B$, significantly larger.  Deviations of the distribution of the true magnetic moments from these values can be viewed as evidence of the mixing of basis states in the true energy eigenstates.  It will be recalled that these results are for the particular case of total angular momentum equal to twice the atomic angular momentum, $J=2j$.  Other manifolds of states will have analogous statistical distributions, of course.  
  
  \subsection{Magnetic Moments of the Fully Coupled Molecule}
  
  Realistically, the distribution of magnetic moments can be strongly modified by channel couplings in the physical model of the molecule.  
  Having the matrix representation in hand, we can compute the magnetic moments in the model.  Generically, at any value of the magnetic field $B$, suppose the Hamiltonian is written
\begin{eqnarray}
H = H_\mathrm{mol} + {\cal M} B,
\end{eqnarray}
where $H_\mathrm{mol}$ is the complete molecular Hamiltonian in zero field and ${\cal M}$ is a magnetic moment matrix with ${\cal M} B =  H_\mathrm{B}$.  Suppose we desire the magnetic moments at  a magnetic field $B$.  Then we contemplate a perturbation of the field $\Delta B$ and write the Hamiltonian
\begin{eqnarray}
H = H_\mathrm{mol} + {\cal M} B + {\cal M} \Delta B.
\end{eqnarray}
   Let $U$ be the matrix whose columns are the eigenvectors of $H_\mathrm{mol} + {\cal M}B$, so that the  energies of the molecule at field $B$ are the diagonal elements of
\begin{eqnarray}
\mathrm{ diag}(E_{\alpha}(B)) = U^T \left( H_\mathrm{mol} + {\cal M}B \right)U.
\end{eqnarray}
Casting the full Hamiltonian in this basis, we get
\begin{eqnarray}
U^T H U = \mathrm{diag} E_{\alpha}(B) + U^T {\cal M}U \Delta B,
\end{eqnarray}
whereby, in the perturbative limit, the magnetic moments of the states are given by 
\begin{eqnarray}
\mu_{\alpha} \approx  \frac{  \Delta E_{\alpha} }{ \Delta B }
=  \left( U^T {\cal M} U \right)_{\alpha \alpha}.
\end{eqnarray}
This expression is used to calculate the magnetic moments, in the zero-field limit, in the examples below.

\section{Results}

 \subsection{Comparison of the basis states}
 
Given the two standard basis sets, in the body and laboratory frames, the first question is to inquire which, if either, is a better representation of the full energy eigenstates of the molecule.  To this end, we deploy the participation number, defined as follows.   Any eigenstate $| \alpha \rangle$ of the Hamiltonian is expressed in a basis $|i \rangle$ by $|\alpha \rangle = \sum_i |i \rangle \langle i | \alpha \rangle$.  Given this expansion, the participation  number is given by \cite{dAlessio16_AP}
\begin{eqnarray}
D({\alpha}) = \left(  \sum_i |\langle i | \alpha \rangle|^4 \right)^{-1}.
\label{eq:Shannon}
\end{eqnarray}
This and related entropies, such as the Shannon entropy, serve to measure the deviation of the energy eigenstates $|\alpha \rangle$  from the basis states $|i \rangle$ from which they are forged.  For example, if the energy eigenstate is already uniquely identified by the basis state $|i \rangle$, i.e., if $|\langle i | \alpha \rangle| = 1$, then $D(\alpha)=1$; only a single basis state participates.  Alternatively, if $n$ states equally participate and $|\langle i | \alpha \rangle | = 1/\sqrt{n}$ for each of them, then $D(\alpha)=n$ counts them.  In this paper we prefer the participation number  to the Shannon entropy because of the significance of the value $D(\alpha)=1$ in identifying states of good quantum  number.  

\begin{figure}[h]
\includegraphics[width=9cm]{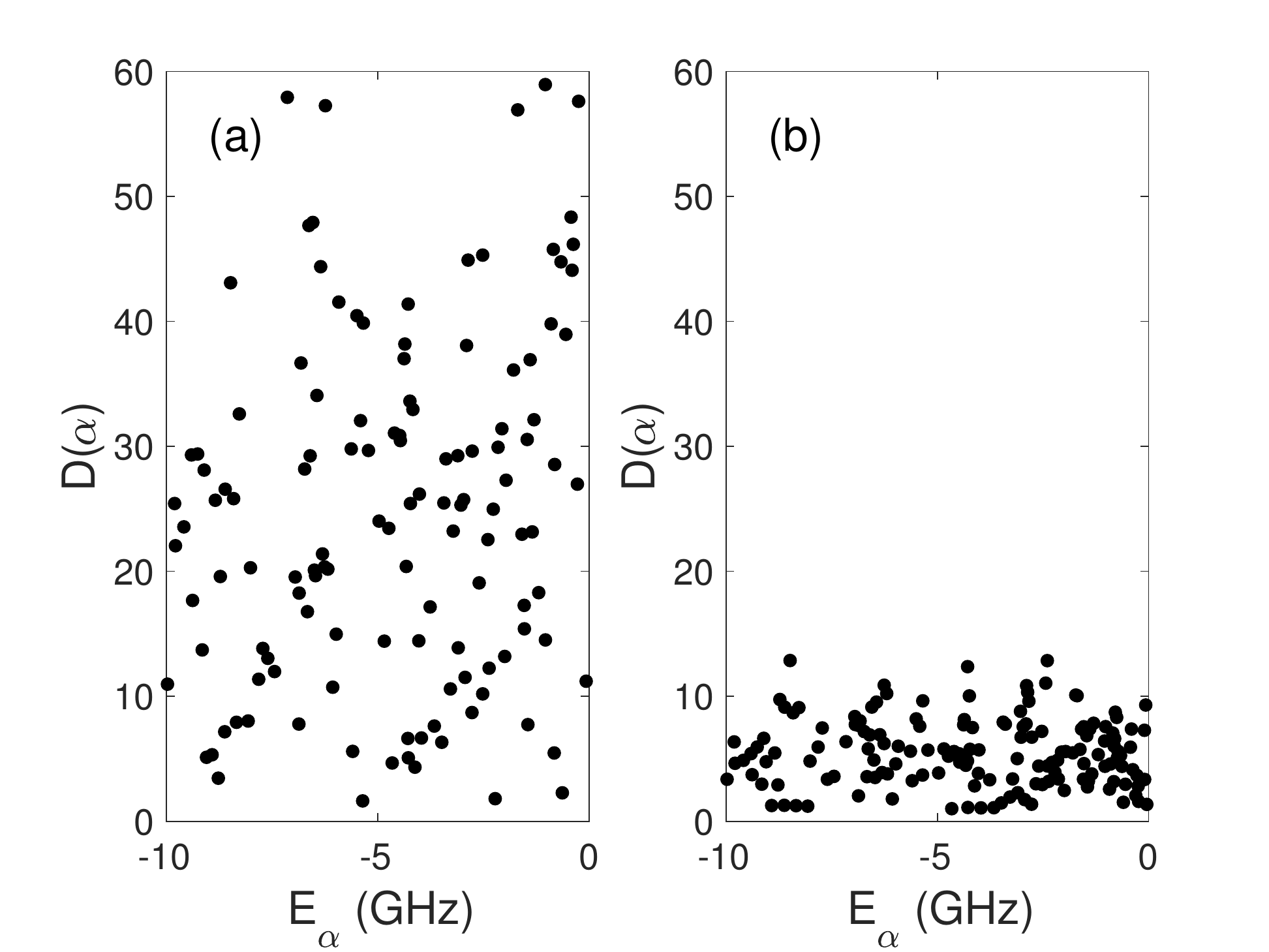}
 \caption{ Participation  number  $D(\alpha)$ of each energy eigenstate $|\alpha \rangle$ in the model versus the energy $E_{\alpha}$ of the  state.  $D(\alpha)$ is computed with respect to the coupled body frame $|j_{12} {\bar \Omega};JM \rangle$ in (a); and with respect to the coupled lab frame basis $|[j_{12}L]JM \rangle$ in (b).   }
 \label{fig:entropy}
\end{figure}

We have calculated an exemplary spectrum, using the model described in the previous section, in terms of both the body and the laboratory basis set, assuming zero magnetic field.  These are converged so as to give the same spectrum for both calculations.  In Figure \ref{fig:entropy} we plot the participation number  of the states versus the energy of the state, for the part of the spectrum lying 10 GHz below the dissociation threshold.  This is shown for both the coupled body frame basis set (a) and the lab frame basis set (b).  

It is immediately clear that $D(\alpha)$ is greater for the body frame basis than for the lab frame basis, thus the latter is more likely a reasonable description of the states.  This comparison affords complementary perspectives on the origin of chaos.  In the body frame, potential interactions such as the model Born-Oppenheimer curves, the dipole-dipole interaction, and the anisotropic van der Waals interaction, are diagonal in the quantum  number ${\bar \Omega}$, which ought to make this quantum  number appropriate for the description of the states.  However, near threshold the molecule, rotating with high angular momentum, is subject to strong Coriolis coupling, which thoroughly mixes the different ${\bar \Omega}$ states.  From this point of view, chaos arises from couplings due to kinetic energy,  

From the other perspective, in the lab frame the kinetic energy is already diagonal in the rotational quantum number $L$.  The states of different $L$ are mixed by the potential interaction terms, primarily the anisotropic van der Waals interaction.  This is a less significant mixing of the basis states, as evidenced by the smaller participation number.  We may therefore try to identify the magnetic moments in terms of the laboratory-frame $g$-factors in Eqn.~(\ref{eq:statistical_lab}). 

\begin{figure}[h]
\includegraphics[width=9cm]{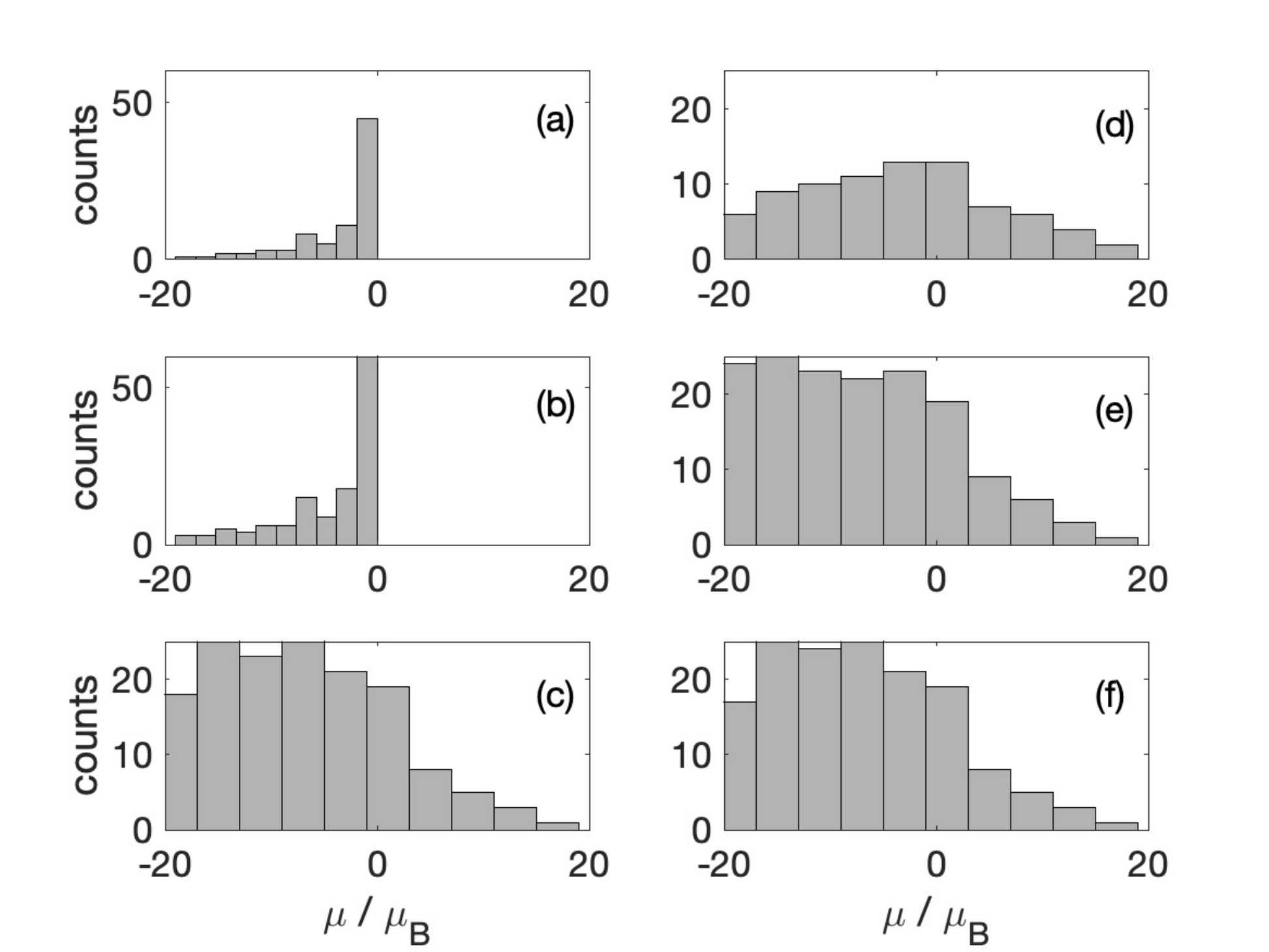}
\caption{Distributions of magnetic moments in the model Dy$_2$ molecules, at various levels of approximation.  Left panels: body frame.  Right panels: lab frame.  Top row: statistical moments.  Middle row: basis moments (see text).  Bottom row: the final, physical moments, which are of course the same in both calculations.  }
\label{fig:moment_distribution}
\end{figure}

\subsection{Magnetic moments in the coupled states}

The superiority of the lab frame over the body frame is shown in more detail by considering the distribution of magnetic moments.  To see this, we present in Fig.~\ref{fig:moment_distribution} various distributions of magnetic moments for the $J=16$, $M=-16$ state of Dy$_2$, computed at various levels of approximation.    In the first column, panels (a), (b), (c), we have results calculated in the body frame basis.  In the second column, panels (d), (e), (f), are results from the lab frame calculation.  In each case, the first row represents the statistical distribution of magnetic moments, as given by Equations (\ref{eq:statistical_body}) and (\ref{eq:statistical_lab}), weighting the occurrence of each possible quantum number equally.  The middle row describes the distribution of ``basis moments,'' those that belong to the ro-vibrational states defined in (\ref{eq:ro_vibrational_basis}).  Finally, the third row of Figure \ref{fig:moment_distribution} includes the physical distribution of magnetic moments, including all channel couplings.  It is of course the same in panels (c) and (f), as the physical result does not care for the basis used to calculate it.  Recall that the moments for states of Dy$_2$ with $J=16$ should lie between $\pm 19.9 \mu_B$.  This entire range is represented in the actual moments, although they are biased toward negative values for the $M=-16$ state considered.  

Panels (a) and (d) represent the statistical moment distributions, which have the means and standard deviations given by  (27,28), respectively.  The body frame moments in (a) are heavily weighted near zero, since the value of ${\bar \Omega}=0$ occurs many times in this set of quantum  numbers, once for each value of $j_{12}$.  The distribution in (a) is far from the physical distribution in (c) since, as noted above, states with different ${\bar \Omega}$ get strongly mixed by Coriolis forces in the real molecule.  By contrast, the statistical distribution of magnetic moments in the lab frame, (d), already  resembles the final distribution in (f).  The lab frame $g$-factors are already a good first guess at the molecular moments, but differ in details.  

The second row adds a little bit to the physics of the molecules, by incorporating the vibrational structure while still assuming rigorously good quantum numbers in either basis.  Because vibrational motion in the diabatic potential energy surfaces is considered, the energies shift somewhat and so do the magnetic moments, from the statistical distribution.  These shifts do not affect the body frame moments much, i.e., panels (a) and (b) are similar.  

In the lab frame, the main difference between the statistical moments and the basis moments is that the basis moments in panel (e) tend to favor lower values than the statistical moments in panel (d).  These basis moments include vibrational motion in $R$, hence are influenced by (among other things) the centrifugal potential $\hbar^2 L(L+1)/2m_r R^2$ in each channel.  For larger values of $L$, the distance between the inner and outer turning points of the diabatic potential $V_d$ are closer together, contributing to higher radial kinetic energy.  As a consequence, the vibrational spacing is larger and there are fewer states of high-$L$ to be found in the energy interval considered.  According to  (\ref{eq:statistical_lab}), these high-$L$ states tend to correspond to lower $g$-factors, or higher magnetic moments for the $M=-16$ states we consider here [see (\ref{eq:g_def})].  Hence, states of higher $L$ are less common in (e) than in (d), with the consequence that there are fewer positive magnetic moments.  

The physical distributions of magnetic moments, including the full channel coupling, are given in panels (c) and (f) for this model.  This calculation includes all the additional off-diagonal coupling between the ro-vibrational states used in panels (b) and (e).  These couplings influence the body frame moments dramatically, and the lab frame results less so.  In either case, however, the distributions in the fully coupled calculations must be the same, as seen by the means and standard deviations of the moments presented in Table I.  Taken together, we conclude that  in the lab frame basis, states with good values of the quantum numbers $j_{12}$ and $L$, distributed as in Fig.~(\ref{fig:moment_distribution}e) already very nearly comprise the correct distribution.

\begin{table}
\centering
\begin{tabular}{| l | r | r |}
\hline
 & body frame & lab frame \\
 \hline
${\bar \mu}$, statistical & -3.49  & -3.49 \\
$\sigma(\mu)$, statistical & 4.35 & 9.32 \\
\hline
${\bar \mu}$, basis & -4.18  & -7.37 \\
$\sigma(\mu)$, basis & 4.88 & 8.64 \\
\hline
${\bar \mu}$, full & -7.23  & -7.23  \\
$\sigma(\mu)$, full & 8.28 & 8.28  \\
\hline
\end{tabular}
\caption{Mean, ${\bar \mu}$, and standard deviation, $\sigma(\mu)$}, for the distributions of magnetic moments shown in Figure \ref{fig:moment_distribution}.  
\end{table}

\subsection{Reduced density matrices}

The laboratory frame is unambiguously the better set of quantum numbers to describe the molecules and their magnetic moments.   The basis is not perfect, however; mixing of these states really does occur.  A further look into the structure of the molecules would investigate which degrees of freedom are most strongly mixed and which are weakest, i.e., which of the several degrees of freedom possesses the best quantum numbers.  

To quantify the goodness of a given quantum number, we employ additional concepts from information theory.  In the first step, we cast the problem in the language of density matrices.  In terms of the expansion coefficients $\langle \alpha | i \rangle$ of the state $| \alpha \rangle$ in the basis $|i\rangle$, we construct a diagonal density matrix with elements
\begin{align}
\rho_{ii^{\prime}}(\alpha) = |\langle \alpha | i \rangle |^2 \delta_{ii^{\prime}}.
\end{align}
This has the essential property that  a density matrix should possess, namely, $\mathrm{Tr}(\rho)=1$.  This $\rho$ would be analogous to a pure state if it had only a single nonzero element, whereby we would have $\mathrm{Tr}(\rho^2)=1$.  More generally, $\mathrm{Tr}(\rho)$ falls short of unity, and the occurrence of multiple basis states in the eigenstate corresponds to the density matrix representing a mixed, as opposed to a pure, state.  This is indeed how one makes the intellectual transition to the entropy, given as
\begin{align}
S(\alpha) = - \mathrm{Tr}( \rho \ln \rho).
\end{align}

Casting the state in terms of this apparent density matrix allows us to extract reduced density matrices for the different degrees of freedom.  For example, the laboratory basis is indexed by its quantum  numbers $|i \rangle = |j_{12},L,v \rangle$ (assuming fixed $J$,$M$).  Then we can extract the reduced density matrix in, say, the $j_{12}$ quantum  number via
\begin{align}
{\bar \rho}_{j_{12}}(\alpha) = \sum_{L,v} \rho_{j_{12},L,v;j_{12},L,v} (\alpha)
= \sum_{L,v} | \langle \alpha | j_{12},L,v \rangle |^2.
\end{align}
Treating $j_{12}$ as the only remaining degree of freedom, we can assign a reduced entropy to the state, or in our case, a reduced participation  number, given by
\begin{align}
{\bar D}_{j_{12}}(\alpha) = \left( \sum_{j_{12}} {\bar \rho}_{j_{12}}^2(\alpha) \right) ^{-1}
= \left( \sum_{j_{12}} \left[ \sum_{L, \nu} | \langle \alpha | j_{12}, L, \nu \rangle |^2 \right]^2 \right) ^{-1}
\end{align}
 Low values of ${\bar D}_{j_{12}}(\alpha)$ correspond to states where $j_{12}$ is a nearly good quantum  number in state $| \alpha \rangle$, regardless of whether the other quantum  numbers are good or not.  The analogous reduced density matrices and participation numbers ${\bar D}_{L}(\alpha)$, ${\bar D}_{v}(\alpha)$  for the other degrees of freedom can be defined analogously.

\begin{figure}[h]
\includegraphics[width=8cm]{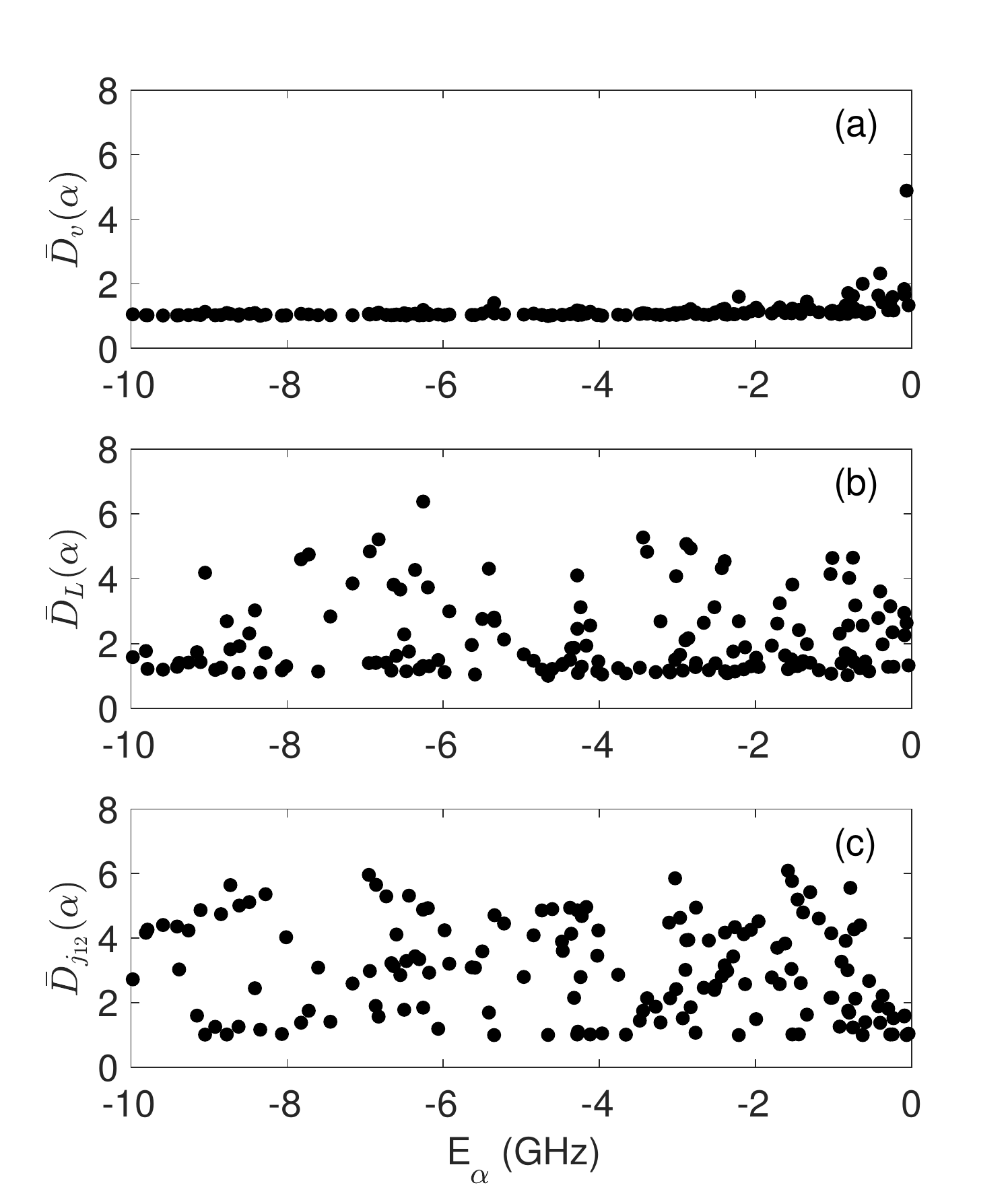}
\caption{Reduced participation numbers for the three relevant quantum  numbers in the lab-frame basis set, for the $J=16$, $M=-16$ states of Dy$_2$ near threshold.  }
\label{fig:participatoin_vs_energy}
\end{figure}

    The participation number is shown for the three relevant quantum numbers of the lab frame basis set in Figure \ref{fig:participatoin_vs_energy}, for the same data as in Fig.~\ref{fig:moment_distribution}.  Panel (a) shows ${\bar D}_v(\alpha)$ for the vibrational quantum number of the atoms.  It is almost uniformly equal to unity, except perhaps very near the dissociation threshold.  We conclude that vibrational states are only weakly mixed, in this basis, upon the introduction of potential coupling between the channels.  
    
    Figure (\ref{fig:participatoin_vs_energy}b) shows the reduced participation number ${\bar D}_L(\alpha)$ for the rotation of the atoms about their center of mass.  We note that for the current model with $J=16$, $L$ can take all the even values up to $32$, or $7$ values in all, whereby ${\bar D}_L(\alpha)$ could conceivably be as large as $7$, for thorough mixing of all the $L$ states. While ${\bar D}_L(\alpha)$ occasionally approaches this limit for some states, nevertheless it is nearly equal to unity for a large fraction of the states in this energy range.  This is consistent with the tale told above, that $L$ in the laboratory frame basis set is appropriate for describing the states.  The fact that ${\bar D}_L(\alpha)$ is often close to one is evidence that the states are not thoroughly chaotic, as they do not strongly mix the different $L$ states.  It is significant, however, that some states apparently do mix various $L$ states.
    
    The real mixing of basis states occurs for the total spin angular momentum $j_{12}$, whose reduced participation number ${\bar D}_{j_{12}}(\alpha)$ is shown in (\ref{fig:participatoin_vs_energy}c).  For the Dy$_2$ model considered, $j_{12}$ can take even values form 0 to $2j=16$, or nine values in all,  setting an upper limit to the value of ${\bar D}_{j_{12}}(\alpha)$.  This limit is never quite achieved for these states, but there is certainly more scatter in the values of ${\bar D}_{j_{12}}(\alpha)$ than there is for ${\bar D}_L(\alpha)$.  It appears, therefore, that the greatest channel mixing that contributes to the chaotic behavior of the molecule lies in the mixing of the atomic spins.  Nevertheless, even in this case there exist states with good values of $j_{12}$, where ${\bar D}_{j_{12}}(\alpha) \approx 1$.  
    
    \subsection{Regularities of the eigenstates}
   
\begin{figure}[h]
\includegraphics[width=8cm]{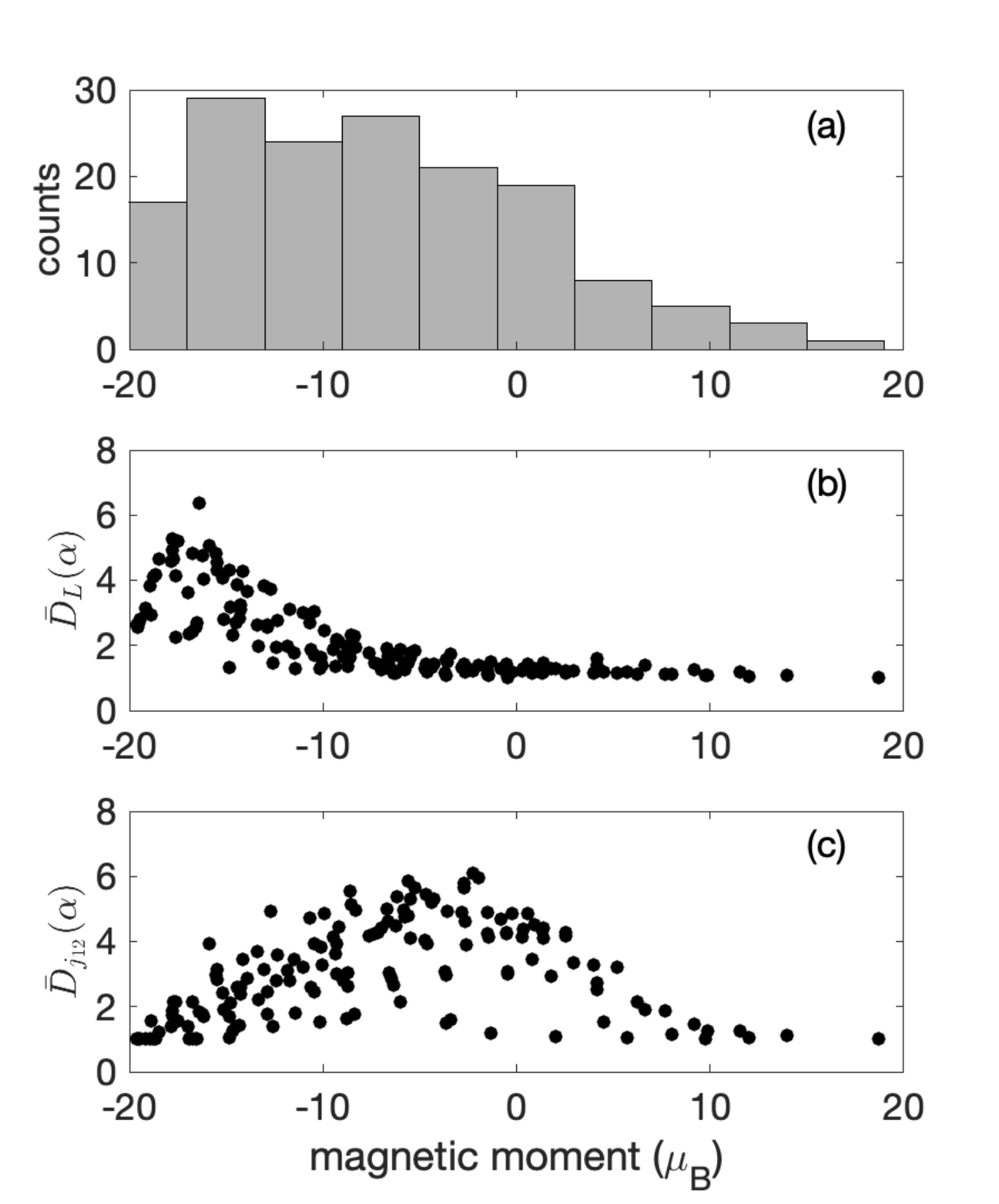}
\caption{Reduced participation numbers for $j_{12}$ and $L$, for the $J=16$, $M=-16$ states of Dy$_2$ near threshold.  These quantities are plotted versus the magnetic moment of the states, also counted in the histogram in panel (a).  }
\label{fig:participation_vs_moment}
\end{figure}

It is instructive to plot the participation numbers for the angular momentum degrees of freedom in an alternative way, as in Figure \ref{fig:participation_vs_moment}.  Panels (b) and (c) show, respectively, the reduced participation numbers ${\bar D}_L(\alpha)$ and ${\bar D}_{j_{12}}(\alpha)$ for each eigenstates, as a function the magnetic moment of that state.  As a reference, panel (a) repeats the histogram of the magnetic moment distribution from Fig.~\ref{fig:moment_distribution}.

In this figure a semblance of order emerges in the correlation between participation number and magnetic moment.  Namely, as shown in  Fig.~\ref{fig:participation_vs_moment}(b), states with  the highest magnetic moments, down at least to the mean ${\bar \mu} =  -7.23 \mu_B$, have low participation number ${\bar D}_L(\alpha)$ -- they are states where $L$ is a good quantum  number.  The states with lower magnetic moments are more often mixtures of states with different $L$-values.  Likewise, a clear trend emerges in Fig.~\ref{fig:participation_vs_moment}(c).  States with extreme values of $\mu$, either high near $20 \mu_B$ or low near $-20 \mu_B$, tend to contain few $j_{12}$ values.  By contrast, states with  intermediate values, around the mean ${\bar \mu}$, mix together several $j_{12}$ states.

\begin{figure}
\includegraphics[width=9cm]{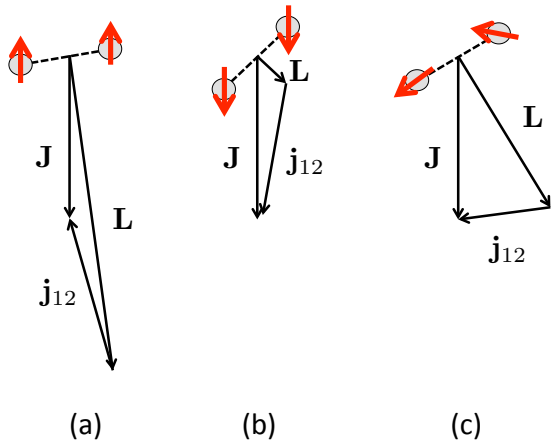}
\caption{Schematic, classical representation of the angular momentum coupling of the molecule.   Grey circles represent the atoms, thick red arrows are their  magnetic moments.  Shown are the cases: (a), large values of $L$; (b), small values of $L$; (c) intermediate values of $L$.}
\label{fig:sketchy}
\end{figure}

These results are qualitatively explained using semiclassical angular momentum coupling diagrams \cite{Zare}, as in Figure \ref{fig:sketchy}.  In this diagram the ${\hat z}$ direction of the laboratory axis points upward, whereby the angular momentum state  $J=16$, $M=-16$ is represented by the arrows pointing straight down in all figures.  (In the semiclassical representation this vector would make a small angle $\cos^{-1}(16/ \sqrt{ (16 \times 17) })$ with respect to the axis, which we here disregard for clarity of the diagram.)  In the laboratory frame, this large angular momentum is described as the sum of the rotation ${\bf L}$ and the total spin ${\bf j}_{12}$, indicated in the figure for positive, negative, and near-zero values of magnetic moment in (a), (b), (c), respectively.  In each case, these angular momenta determine the relative orientation of the atoms (grey circles), and indicate the kind of orientation of the spins (thick red arrows).  In this picture, the more up the spins are allowed to point, the greater their magnetic moment.  

Consider first the case of large positive magnetic moment Fig.~\ref{fig:sketchy}(a).  To achieve this result, the  spin angular momentum is near its maximum value $j_{12}=16$. There are few comparable $j_{12}$ states available to mix together, so in this limit so $j_{12}$ remains a reasonably good quantum  number.  In order to attain the total angular momentum ${\bf J}$, the rotational angular momentum must be near its maximum value $L=32$, but pointing in the other direction.  Consequently, it is not mixed with many other $L$-values either, and both $j_{12}$ and $L$ remain good quantum numbers for large magnetic moment.

Next consider large negative magnetic moment, Fig.~\ref{fig:sketchy}(b).  To achieve this result, $j_{12}$ must again be near its maximum magnitude, but with classical vector pointing downward, so $j_{12}$ remains a nearly good quantum  number.  In this case, however, $L$ must be small in magnitude, and  can take various small values while requiring only small changes in ${\bf j}_{12}$ to add to the total $J$.  Even small values of ${\bf L}$ can represent states where the molecular orientation makes a significant angle with respect to the spin axis, as indicated in the figure.   Therefore, in the presence of the kind of spin-rotation coupling occasioned by the anisotropic van der Waals interaction, the available rotational states can be mixed together.  While $j_{12}$ is a nearly good quantum number, $L$ is not.  

For intermediate magnetic moments, the vector diagram is more like Fig.~\ref{fig:sketchy}(c).  Here $j_{12}$ is free to run over many intermediate values, which can be composed of many different orientations of the individual spins ${\bf j}_1$ and ${\bf j}_2$, which can then interact strongly and anisotropically as the molecules rotates.  Here is where the primary channel mixing occurs: at intermediate-sized  magnetic moment.  

In summary, not only are the various energy eigenstates of the molecule not all chaotic, in the sense of strongly mixing eigenstates, but moreover the states that do show this strong mixing are empirically identifiable via trends in their intermediate magnetic moments.  

\section{Conclusion}

Every chaotic system, governed by random matrix theory, is chaotic in the same manner.  But each system that is only partially chaotic experiences chaos in its own way.  Here we have explored the zero-magnetic-field spectrum of a set of lanthanide dimer van der Waals states, to locate where their chaos resides. Among the $|JM \rangle = |16-16 \rangle$ states considered for Dy$_2$,  we find that states belong to one of three realms of qualitatively different chaoticity, loosely correlated  to the magnetic moment ${ \mu}$ of the state.  States with the highest values of $\mu$ tend to be non-chaotic and described by the quantum  numbers $j_{12}$ and $L$; states of the lowest  $\mu$ are mildly chaotic due to mixing of the orientation of the molecular rotation $L$; and states with intermediate values of $\mu$ near the mean are ``just right'' for chaos, capable of mixing both the spin and rotation states.  

This report has dealt only with molecules associated with the spin-stretched atomic states $|jm \rangle = |8-8 \rangle$ most closely allied with experiment, but many other manifolds of states exist.  Future work should be able to find similar systematics in the spectra and establish a zoology of van der Waals lanthanide dimers.  More significantly, the results remain to be extended to the case of nonzero magnetic field.  One presumes the appearance and pattern of chaos may take different forms when states of different angular momentum $J$ are coupled and the molecules become overall more chaotic \cite{Maier15_PRX,Makrides18_Sci}.  In this context it is worth noting that even more exotic states of lanthanide dimers have been proposed, which possess large electric dipole moments as well as large magnetic dipole moments, providing additional opportunities for introducing and probing chaos \cite{Lepers18_PRL,Li19_PRA}.

In the broader sense, these results imply the ability to identify molecular states with qualitatively different manifestations of chaos by virtue of their magnetic moment.  This ability can be useful in dynamical studies of these chaotic molecules.  For example, having prepared the molecule in a particular state, a sudden quench to a different magnetic field value will project this state onto a host of other energy eigenstates and  will initiate dynamics.  Knowing what the states are likely to be like, one can imagine different quenches to and from molecules that are either rotationally chaotic, spin-chatoic, or both.  The richness of the resulting dynamics remains to be contemplated.

\section*{Acknowledgements}  
We gratefully acknowledge useful discussions with M. Lepers.  This material is based upon work supported by the National Science Foundation under Grant Number PHY 1734006 and Grant Number PHY 1806971.  L.D.A. acknowledges the financial support of the Czech Science Foundation (Grant No. 18-00918S).


\begin{thebibliography}{20}
\bibitem{mean_note} Note that this mean value is given by the average of the diagonal elements of the operator ${\cal M}$, that is, $(1/N)\sum_{i=1}^N \langle i | {\cal M} | i \rangle = (1/N)\mathrm{Tr}({\cal M})$, and the trace of the operator is basis-independent.
\end{thebibliography}
\end{document}